\documentstyle[prd,preprint,aps,epsfig]{revtex}

\draft

\author{S. Hegyi}

\title{Asymptotic multiplicity scaling: a renormalization group perspective}

\address{KFKI Research Institute for Particle and Nuclear Physics \\ of the
	Hungarian Academy of Sciences, \\ H-1525 Budapest 114, P.O. Box 49.
	Hungary}

\date{\today}

\begin{document}

\pagestyle{plain}
\pagenumbering{arabic}

\maketitle

\begin{abstract}
 A generalization of the Polyakov-Koba-Nielsen-Olesen scaling law of the
 multiplicity distributions $P(n,s)$ is developed. It states that a suitable
 change in the normalization point of $P(n,s)$ compensated by a rescaling
 can restore data collapsing onto a universal curve if the original scaling
 rule is violated. We show that the iteratively
 executed transformation of $P(n,s)$ can be viewed as varying the collision
 energy. The $e^+e^-$ and $p\bar p$ multiplicity data at top energies are
 found to exhibit a fixed point property of the iteration.
\end{abstract}


This year is the 25th anniversary year of the influential work of
Koba, Nielsen and Olesen concerning the asymptotic scaling behavior of
multiplicity distributions~\cite{KNO}. They put forward the hypothesis
that at ``sufficiently high'' energies $s$ the probability distributions
$P(n,s)$ of detecting $n$ final-state hadrons in a certain collision
process should exhibit the scaling relation
\begin{equation}
        P(n,s)=\frac{1}{\langle n(s)\rangle}\,
        \psi\!\left(\frac{n}{\langle n(s)\rangle}\right)
        \label{Scal_0}
\end{equation}
where $\langle n(s)\rangle$ is the average multiplicity at collision
energy $s$. According to Eq.~(\ref{Scal_0}), $P(n,s)$ is a homogeneous
function of degree $-1$ of $n$ and $\langle n(s)\rangle$. The
homogeneity rule states that the multiplicity distributions are simple
rescaled copies of the universal function $\psi$, i.e., the change
of collision energy $s$ amounts only to a change of scale in the shape of
$P(n,s)$. The above prediction caused immediately a great deal of activity
in the experimental and theoretical analysis of multiplicity distributions
which remained immense during the past 25~years. It should be mentioned
here that Polyakov arrived at the scaling law  Eq.~(\ref{Scal_0}) already
in 1970 postulating a similarity hypothesis for strong interactions in
$e^+e^-\to\mbox{hadrons}$ annihilation~\cite{POL}.
Despite of this fact, the phenomenon
became known as KNO scaling in the high-energy physics community.

The main goal of the present Letter is to extend the domain of data
collapsing behavior of $P(n,s)$ beyond the scaling
relation Eq.~(\ref{Scal_0}). In particular, we propose an iterative
procedure capable of finding homogeneity rules for $P(n,s)$ if the
scaling hypothesis Eq.~(\ref{Scal_0}) is violated. We shall consider in
detail the fixed points of the iteration, their domains of attraction
and the constraints they put
on the asymptotic scaling form of $P(n,s)$. Our approach is
inspired by the renormalization group methods in the theory of
critical phenomena. In the beginning of the paper we will
review some basic results of Ref.~\cite{HES} which should be considered
as the first part of the present work; it is cited as Paper~1
in the followings.

We start with some elementary properties of the scaling function $\psi(z)$,
$z$ denoting the scaled multiplicity
$n/\langle n(s)\rangle$. Finite energy discreteness effects at moderate
$\langle n(s)\rangle$ are usually taken into account via Poisson transform
which is discussed in Paper~1. The continuous probability density $\psi(z)$
fulfills the normalization conditions
$\int_0^\infty\psi(z)dz=\int_0^\infty z\,\psi(z)dz=1$. The latter one,
yielding the constraint $\langle z\rangle=1$, defines a
second properly normalized scaling function:
$\psi^\prime(z)=z\,\psi(z)=nP(z)$. Obviously, the moments
of $\psi(z)$ coincide with the normalized moments of $P(n,s)$,
\begin{equation}
	{\cal C}_q=\int_0^\infty z^q\psi(z)dz=
	\frac{\langle n^q(s)\rangle}{\langle n(s)\rangle^q}.
	\label{C_q}
\end{equation}
They are independent of collision energy $s$ if the scaling hypothesis
Eq.~(\ref{Scal_0}) holds valid. The moments
of $\psi^\prime(z)$ are given by
${\cal C}_q^\prime=\int_0^\infty z^q\,\psi^\prime(z)dz={\cal C}_{q+1}$
i.e. the difference between the two scaling functions amounts
to a shift in the ranks of their moments.

Let us now recapitulate a scaling argument first presented in
Paper~1. Assume that violation of the original scaling hypothesis
Eq.~(\ref{Scal_0}) is observed and
we measure energy dependent ``scaling'' functions
$\psi(z,s)$ and $\psi^\prime(z,s)$. A key point to the
later developments is the
observation that $\psi^\prime(z,s)$ enables one to perform a rescaling
of type Eq.~(\ref{Scal_0}) yet again to arrive at data collapsing
behavior. The modified scaling hypothesis reads
\begin{equation}
	\psi^\prime(z,s)=\frac{1}{{\cal C}_1^\prime(s)}\,
	\psi_1\!\left(\frac{z}{{\cal C}_1^\prime(s)}\right)
\end{equation}
where the subscript of the new scaling function refers to the first step
of an iterative sequence of transformations that can be performed
on $P(n,s)$. For quantities corresponding to the $0$th step
the subscript will be omitted. Our new scaling
variable is $z_1=z/{\cal C}_1^\prime(s)=z/{\cal C}_2(s)$
and the first-iterate  moments
${\cal C}_{q,1}=\int_0^\infty z_1^{\,q}\,\psi_1(z_1)dz_1$
are related to the original ones through
\begin{equation}
	{\cal C}_{q,1}=\frac{{\cal C}_q^\prime(s)}
	{[{\cal C}_1^\prime(s)]^q}=
	\frac{{\cal C}_{q+1}(s)}{[{\cal C}_2(s)]^q}.
	\label{C_q1}
\end{equation}
Expressing $\psi_1(z_1)$ in a similar manner we get
$\psi_1(z_1)={\cal C}_2(s)\,n\,P[z/{\cal C}_2(s)]$
which can be rewritten according to
\begin{equation}
	\psi_1(z_1)=\frac{\langle n^2(s)\rangle}{\langle n(s)\rangle}\,
	\frac{n}{\langle n(s)\rangle}\,
	P\!\left(n\left/\frac{\langle n^2(s)\rangle}{\langle n(s)\rangle}
	\right.\right).
	\label{psi_1}
\end{equation}
It is seen that $\psi_1(z_1)$ obeys the same structure as the original
scaling function $\psi(z)$, namely,
$\psi_1(z_1)=\langle n(s)\rangle_1\cdot P_1[n/\langle n(s)\rangle_1]$.
The first-iterate multiplicity distribution $P_1(n,s)$ and its moments
$\langle n^q(s)\rangle_1$ are obtained by changing the normalization point
of $P(n,s)$ from the $0$th moment to $\langle n(s)\rangle$ and rescaling by
$\langle n(s)\rangle$ to maintain the overall normalization. The
distribution $P_1(n,s)$ is known in the mathematical literature as the
first order moment distribution of $P(n,s)$.

If $\psi_1(z_1)$ is not independent of collision energy $s$, the
transformation rule described above can be repeated iteratively until the
appearance of data collapsing onto a universal scaling curve. The details
are presented in Paper~1, here we recall only the final result. In the $i$th
step of the iteration the connection between $P_i(n,s)$ and the original
multiplicity distribution $P(n,s)$ is provided by
\begin{equation}
	P_i(n,s)=\frac{n^i}{\langle n^i(s)\rangle}\,P(n,s)
	\label{P_k}
\end{equation}
i.e. the normalization point of $P(n,s)$ is changed to
$\langle n^i(s)\rangle$ and a rescaling by $\langle n^i(s)\rangle$ is made
to preserve proper normalization. Thus the necessary condition of
performing the $i$th iteration step is the existence of the moments of
$P(n,s)$ up to $i$th order. $P_i(n,s)$ is the moment distribution
of order~$i$ of $P(n,s)$. Data collapsing of
$P_i(n,s)$ onto the scaling function
$\psi_i(z_i)=\langle n(s)\rangle_i\cdot P_i[n/\langle n(s)\rangle_i]$
occurs when the normalized moments at the previous iteration step exhibit
the ``monofractal'' type behavior
${\cal C}_{q,i-1}(s)\propto[{\cal C}_{2,i-1}(s)]^{q-1}$
with constants of proportionality independent of collision energy $s$,
see Eq.~(\ref{C_q1}). It is worth mentioning here that for discrete
probability laws such as $P(n,s)$ the factorial moment distributions,
involving factorial powers of $n$,
arise more naturally as discussed in Paper~1.

It is also argued in Paper~1
that the iterative procedure acting on $P(n,s)$
bears some similarity with the renormalization group (RG) transformations.
They correspond in many important applications to a change in the norm
of the parameters characterizing a physical system~\cite{MA}.
On the one hand,
Eq.~(\ref{P_k}) can be viewed as a Gell-Mann - Low type relationship
in which a multiplicative transformation (here multiplication
by $n^i$) is compensated by a rescaling and a suitable change in the
renormalized parameters (here in the moments). Since the operation
$P_i(n,s)\to P_{i-1}(n,s)$ does not exist the iteration steps constitute
a semigroup. In statistical physics the Kadanoff - Wilson type RG
transformations perform a systematic reduction in the number of degrees
of freedom via e.g. spin decimation. This makes possible to eliminate the
small-scale fluctuations from the problem which are irrelevant to
critical point behavior such as the homogeneity of
thermodynamic functions~\cite{MA}.
In the iterative procedure described above the elimination of small-scale
fluctuations corresponds to the elimination of the
low-order moments of $P(n,s)$ via moment-shifting and rescaling until one
observes a homogeneity rule of type Eq.~(\ref{Scal_0}) for $P_i(n,s)$.

The suggested analogy between the transformation
Eq.~(\ref{P_k}) of multiplicity distributions and the RG methods in
field-theory and statistical physics may seem
too remote at first glance. But there
is a very close relationship between RG ideas and certain concepts
of probability theory~\cite{GJL,YAGS} which naturally fits into our
approach to asymptotic multiplicity scaling. Let us demonstrate it by
considering some properties of Eq.~(\ref{P_k}) in more detail.

{\it Property \#1: Form-invariance under size-biasing}

If we view the Gell-Mann - Low or Kadanoff - Wilson version of the RG
as a transformation acting
on a probability distribution, it is a transformation
that does not change the form of the distribution~\cite{GJL}.
Weighting a probability law according to Eq.~(\ref{P_k}) is known in
statistics as size-biasing of order~$i$. Members of the log-exponential
family of distributions, such as the beta, gamma,
Pearson type~V, Pareto and log-normal to mention but a few,
are form-invariant under size-biasing, i.e. they retain
functional form and only their parameters are affected~\cite{PO}.
Therefore if $P(n,s)$ belongs to the log-exponential
family, our iterative procedure is in accordance with this
particular aspect of
the Gell-Mann - Low and Kadanoff~-~Wilson type RG transformations.

{\it Property \#2: Fixed points and automodel distributions}

A probability distribution is called a scaling- or automodel distribution
if it is invariant under the action of the RG, or, in other words,
if it is a fixed point of the RG transformation~\cite{YAGS}. In our case
automodel distributions are those members of the log-exponential family
which are affected by size-biasing through a scale-change.
A theorem of Ref.~\cite{VSL} states that size-biasing amounts to a
change of scale in the original distribution if its normalized moments
have the form ${\cal C}_q={\,\cal C}_2^{q(q-1)/2}$. This is a well-known
property of the log-normal law whose parameters are transformed by
Eq.~(\ref{P_k}) according to $\nu\to\nu+i\sigma^2$ with $\nu$ and
$\sigma^2$ being the mean and variance of the unbiased distribution.
A log-normal $P(n,s)$ displays fixed point behavior in the following
manner: although the transformation
Eq.~(\ref{P_k}) changes its first moment, this change is scaled out
by constructing $\psi_i(z_i)$ and one arrives at an unaffected
scaling function at each iteration step.
Therefore a log-normally shaped $\psi(z)$ is a fixed point of the
iterative procedure. Let us emphasize that the log-normal
law is not a unique fixed point of the iteration
because the distribution is not uniquely determined by its moments.

{\it Property \#3: Domains of attraction of the fixed points}

Besides finding the fixed points of the RG transformation, it is important
to specify their domains of attraction. A domain of attraction is
the set of initial
probability laws which converge to a given automodel
distribution under the action of the RG transformation~\cite{YAGS}.
The domains of attraction
are analogous to the universality classes of critical phenomena.
Here we are interested in those probability laws which converge to the
log-normal fixed point under the action of Eq.~(\ref{P_k}). Let us choose
$\psi(z)$ to be the generalized gamma density
\begin{equation}
        \psi(z)=\frac{\mu}{\Gamma(k)}\,\lambda^{\mu k}
        z^{\mu k-1}\exp\left(-[\lambda z]^\mu\right)
	\label{GG}
\end{equation}
with shape parameter $k>0$, scaling exponent $\mu>0$ and scale parameter
$\lambda$ restricted to $\lambda=\Gamma(k+1/\mu)/\Gamma(k)$
by the normalization condition $\langle z\rangle=1$.
In the limit $\mu\to0$ and
$k\to\infty$ the scaling function given by Eq.~(\ref{GG}) converges to a
log-normal $\psi(z)$ of variance $\sigma^2$ in such a way that
$k\mu^2\to1/\sigma^2$~\cite{FER}. The generalized gamma distribution is a
member of the log-exponential family and thus it is form-invariant under
size-biasing. The shape parameter $k$ is
changed by Eq.~(\ref{P_k}) according to $k\to k+i/\mu$
whereas the scaling
exponent $\mu$ remains unchanged~\cite{DIA}. The $\mu\to0$, $k\to\infty$
log-normal limit can be achieved by requireing ${\cal C}_{2,i}={\cal C}_2$
in the course of the iteration; the increase of $k$ is compensated by
a decrease of $\mu$ so that the second moment of $\psi(z)$ is not
affected by the variation of the parameters.
Thus suitably standardized generalized gamma
distributions constitute a domain of
attraction of the log-normal fixed point of the iteration.

{\it Property \#4: An inequality for the first moments}

The transformation rule  Eq.~(\ref{P_k}) is a special case of weighting a
probability distribution $f(x)$ with the non-negative weight-function
$w(x)$ according to $w(x)f(x)/\langle w(x)\rangle$. The first moment of the
weighted distribution is greater or smaller than the first moment of $f(x)$
depending on whether $w(x)$ is a monotonously increasing or decreasing
function of $x$. In the case $w(x)=x$, i.e. for size-biasing of order one,
we have the inequality $\langle x\rangle_1>\langle x\rangle$
and in the general case $\langle x\rangle_{i+1}>\langle x\rangle_i$.

{\it Consequences}

In many applications of renormalization group methods the
transformations establish correspondences between physically different
states of the same system. For example, in the Ising model the iterative
repetition of the RG transformation can be viewed
as varying the temperature.
Similar correspondence can be established for asymptotic multiplicity
scaling. According to the above inequality
of the first moments, the iterative sequence of normalization point
changing transformations acting on $P(n,s)$ can be viewed as increasing
the collison energy $s$. The asymptotic scaling relation
Eq.~(\ref{Scal_0}), which states
that the increase of~$s$ amounts to a change
of scale in the shape of $P(n,s)$, is therefore a fixed point property.
An important consequence of our reasoning is the fact that the asymptotic
scaling function $\psi(z)$ can not be arbitrarily shaped:
the above fixed point behavior is satisfied only by those probability
laws which are equivalent in their moments to the log-normal distribution.

In the light of the previous findings it is of interest to estimate the
degree of deviation between the log-normal distribution and the shape of
$\psi(z)$ at asymptotic energies. Making use of the generalized gamma
density~Eq.~(\ref{GG}) the parameter pair ($k,\mu$) is well suited to
measure the departure from fixed point behavior corresponding to the
$\mu\to0$, $k\to\infty$ limit. The shape of the asymptotic $\psi(z)$ can
be reconstructed by fitting the Poisson transform of the generalized
gamma distribution, let us call it HNBD for short, to the experimental
data for $P(n,s)$ available at the highest collision energies~$s$.
The HNBD was invented and developed in Refs.~\cite{HES1,HES2} by the
present author. The analytic form of $P(n,s)$ can be expressed in terms
of generalized special functions:
\begin{equation}
   P(n,s)=
   {\cal N}\;{\sf H}^{1,1}_{1,1}
   \left[
      \,\frac{1}{\theta}\,\left|
      \begin{array}{c}
         (1,\, 1)             \\
         ({\cal K},\, 1/\mu)
      \end{array}
   \right]\right.
   \qquad\mbox{for }0<\mu<1
	\label{lt1}
\end{equation}
\begin{equation}
        P(n,s)={\cal N}\;\Gamma({\cal K})
        \,(1+\theta)^{-{\cal K}}\qquad\mbox{for $\mu=1$ \ (NBD)}
	\label{nbd}
\end{equation}
\begin{equation}
   P(n,s)=
   {\cal N}\;{\sf{}_1{\Psi}_0}
   \left[\left.
      \begin{array}{c}
         ({\cal K},\, 1/\mu)         \\
         -\!\!-
      \end{array}\right|-\theta\,
      \right]
	   \quad\mbox{\ for }\mu>1
	\label{gt1}
\end{equation}
where ${\cal K}=k+n/\mu$, ${\cal N}^{-1}=n!\,\Gamma(k)\,\theta^{-n}$ and
$\theta=\langle n(s)\rangle\Gamma(k)/\Gamma(k+1/\mu)$; of course the shape
parameter $k$ and scaling exponent $\mu$ may also depend on~$s$.
The functions
${\sf H}^{1,1}_{1,1}(\cdot)$ and ${\sf{}_1{\Psi}_0}(\cdot)$ are
particular cases of the Fox- and Wright hypergeometric functions,
respectively~\cite{MS}. The negative binomial distribution
given by Eq.~(\ref{nbd})
is the $\mu=1$ marginal case of Eq.~(\ref{lt1}) for
$\langle n(s)\rangle>k$ and of Eq.~(\ref{gt1})
for $\langle n(s)\rangle<k$.
The Poisson transformed log-normal limit ($\mu=0$) lacks a representation
in terms of known functions.

We have investigated two full phase-space data sets for $P(n,s)$: the
Delphi data at \hbox{$\sqrt s=91$~GeV} in $e^+e^-$ annihilations~\cite{DEL}
and the UA5 data at \hbox{$\sqrt s=900$~GeV}
in $p\bar p$ collisions~\cite{UA5}. For each data set we have performed
several HNBD fits with different scaling exponents~$\mu$. The value of
$\mu$ was varied
in the interval $0.1\leq\mu\leq2$ by stepsize $\Delta\mu=0.1$.
It is worth considering how the $\chi^2$ and the best-fit shape
parameter~$k$ depend on $\mu$. The trends are displayed in Fig.~1.
The top right inset shows the variation of fit quality.
As is seen the $\chi^2$
decreases monotonously towards $\mu=0$, approximately as an
exponential, for both data sets.
The shape parameter~$k$ increases according to a
power-law as $\mu\to0$ with slopes being the same for the two reactions
(the estimated errors of $k$ are too small to be seen). Let us recall that
the convergence to a log-normal law of variance~$\sigma^2$ is such that
$\mu\to0$ and $k\to\infty$ with $k\mu^2\to1/\sigma^2$.
Thus the $\mu$-dependence of $k$ allows us to
estimate the value of $\sigma$ by fitting $k=(\sigma\mu)^{-2}$ to the
data points. The fits are represented by the straight lines in Fig.~1.
The quality of fits and the best-fit value of~$\sigma$ are shown in the
first row of Table~I for each data set. The second row quotes the same
numbers obtained in~\cite{HES2} by fitting the Poisson transform of the
log-normal distribution to $P(n,s)$.
All these results suggest the same conclusion: the asymptotic scaling
function $\psi(z)$, as can be guessed from pre-asymptotic multiplicity data,
is log-normally shaped both in $e^+e^-$ annihilations and in
$p\bar p$ collisions. Interestingly, this is just
the fixed point behavior we would expect
on the basis of the RG approach to asymptotic multiplicity scaling.

Summarizing our results,
we have developed an iterative procedure well suited to find homogeneity
rules of type Eq.~(\ref{Scal_0}) for the multiplicity distributions
$P(n,s)$. The quoted scaling law states that the energy dependence of
$P(n,s)$ is due entirely to the energy dependence of its first moment
$\langle n(s)\rangle$. Thus rescaling $P(n,s)$
by the average multiplicity according to
Eq.~(\ref{Scal_0}) a universal scaling function emerges whose shape is
independent of~$s$. This scaling argument can be extended to the more
general case when the $s$-dependence of $P(n,s)$ can not be attributed
to $\langle n(s)\rangle$ exclusively. To arrive at data collapsing onto
a universal scaling curve, a modified scaling transformation is needed
capable of eliminating the $s$-dependence
of higher-order moments. An obvious solution is to ``shift out''
the moments of the multiplicity distributions up to the required order.
This can be achieved by changing the normalization point of
$P(n,s)$ from the $0$th moment
to $\langle n(s)\rangle$, rescaling by $\langle n(s)\rangle$
to maintain the overall normalization and repeating the two-step
transformation until the appearance of the scaling law Eq.~(\ref{Scal_0})
for the iterated distribution. In other words, we have the possibility
that the increase of collision energy~$s$ amounts to a change of scale
not in $P(n,s)$, rather, in a moment distribution $P_i(n,s)$
with $i>0$. Accordingly, the phenomenon of
data collapsing onto a universal scaling curve should be
checked for $i>0$ as well in Eq.~(\ref{P_k}).
During the past 25 years this was traditionally done only for $i=0$.

Besides a whole family of new scaling relations for the multiplicity
distributions, Eq.~(\ref{P_k}) provides a close analogy
with certain RG ideas.
In this respect there is a distinguished role of those probability laws
which are form-invariant under the action of  Eq.~(\ref{P_k}) and the
moment distributions $P_i(n,s)$ are simple rescaled copies of $P(n,s)$.
These distributions are the only ones which may exhibit the
asymptotic scaling behavior Eq.~(\ref{Scal_0}). It is known that all
probability laws having the above property are equivalent in their
moments to the log-normal law. Remarkably, the multiplicity data in
$e^+e^-$ annihilations and in $p\bar p$ collisions at the highest
available energies indicate log-normality of the asymptotic scaling
function $\psi(z)$~---~just as one would expect
on the basis of RG arguments.
To decide whether this is purely accidental (as already happened with
results obtained by related arguments~\cite{ES}) or is an
intrinsic feature of multiparticle production, measurements at even higher
energies are needed. The forthcoming $p\bar p$
data at $\sqrt s=1800$~GeV at Tevatron are awaited with keen interest.

\bigskip\bigskip

This research was supported by the Hungarian Science Foundation
under Grant No. OTKA-T024094/1997.

\bigskip

\begin{table}
\vskip1cm
\begin{tabular}{lcr}
Data set  &  $\sigma$   &   $\chi^2$/d.o.f.  \\ \tableline
$e^+e^-$, $\sqrt s=91$ GeV & $0.199\pm0.001$ & 12.0/18 \\
                           & $0.201\pm0.004$ & 32.4/24 \\
\tableline
$p\bar p$, $\sqrt s=900$ GeV & $0.527\pm0.003$ & 17.3/18 \\
                             & $0.538\pm0.014$ & 32.7/52 \\
\end{tabular}
\bigskip\bigskip
\caption{For each data set the first row displays the result of fitting the
power-law $k=(\sigma\mu)^{-2}$ to the $\mu$-dependence of the HNBD shape
parameter $k$. The fits are represented by the straight lines in Fig.~1.
The second row quotes the outcome of Poisson transformed
log-normal fits to $P(n,s)$, see Ref.~[12] for details.}
\end{table}

\newpage

\begin{figure}
\vskip1cm
\hskip1.7cm\epsfig{file=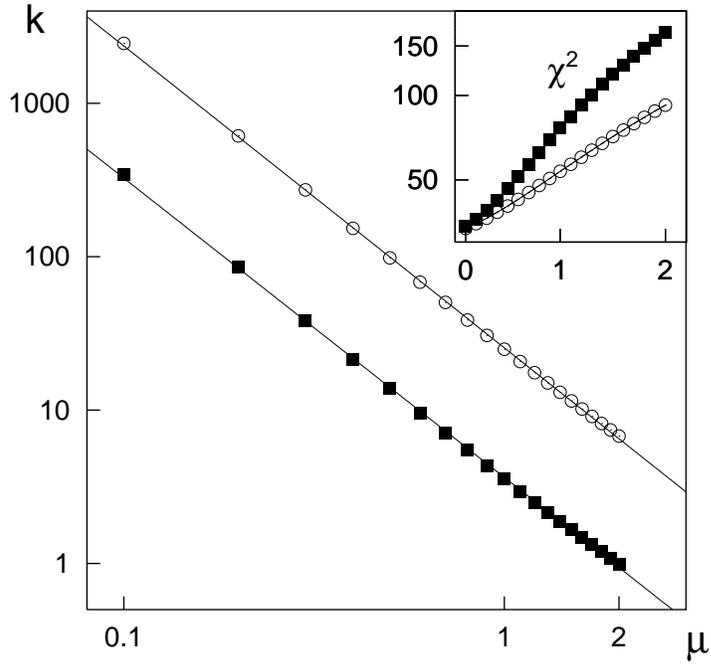}
\bigskip\bigskip
\caption{Variation of the best-fit shape parameter $k$ of the HNBD as a
function of scaling exponent $\mu$ for $e^+e^-$ annihilations at
$\sqrt s=91$~GeV (open circles) and for $p\bar p$ collisions at
$\sqrt s=900$~GeV (solid squares). The straight lines are power-law
fits, see the text and Table~I for details.
The inset in the top right displays
the $\mu$-dependence of $\chi^2$ corresponding to the HNBD fits.}
\end{figure}

\end{document}